# On trees with a maximum proper partial $0-1$ coloring containing a maximum matching

V. V. Mkrtchyan

Department of Informatics and Applied Mathematics, Yerevan State University, Yerevan, 0025, Republic of Armenia, e-mail: vahanmkrtchyan2002@{ysu.am, ipia.sci.am yahoo.com}

*For Anush*

**Abstract**

I prove that in a tree in which the distance between any two endpoints is even, there is a maximum proper partial $0-1$ coloring such that the edges colored by 0 form a maximum matching.

Keywords: Partial proper $0-1$ coloring, Maximum partial proper $0-1$ coloring, Maximum matching

All graphs considered in this paper are finite, undirected and have no loops or multiple edges. $V(G)$ and $E(G)$ denote the sets of vertices and edges of a graph $G$, respectively. The degree of a vertex $x$ in $G$ is denoted by $d_G(x)$. If $X \subseteq E(G)$ then a mapping $f : X \to \{0,1\}$ is referred as a partial $0-1$ coloring of the graph $G$. For $i = 0, 1$ and the partial $0-1$ coloring $f$ of the graph $G$, denote $f_i \equiv \{e \in X / f(e) = i\}$. The partial $0-1$ coloring $f$ is proper if the sets $f_0$ and $f_1$ are matchings of the graph $G$. Denote

$$\lambda(G) \equiv \max\{|f_0| + |f_1| / f \text{ is a proper partial } 0-1 \text{ coloring of the graph } G\}.$$

A proper partial $0-1$ coloring $f$ of the graph $G$ is maximum if $|f_0| + |f_1| = \lambda(G)$. Set:

$\alpha(G) \equiv \max\{|f_i| / i = 0, 1 \text{ and } f \text{ is a maximum proper partial (shortly, MPP) } 0-1 \text{ coloring of the graph } G\}$.

It is clear, that for every graph $G$ $\alpha(G) \leq \beta(G)$, where $\beta(G)$ is the cardinality of a maximum matching of the graph $G$. In this paper I show that if $G$ is a tree in which the distance between any two endpoints is even, the equality $\alpha(G) = \beta(G)$ holds. Non defined terms and conceptions can be found in [1,2].

**Lemma1**. Let $G$ be a graph, $u \in V(G)$, $w \in V(G)$, $(u, w) \in E(G)$, $d_G(u) = 1$. Then there is a MPP $0-1$ coloring $f$ of the graph $G$, such that $|f_0| = \alpha(G)$ and $(u, w) \in f_0$.

**Proof.** Let $f$ be a MPP $0-1$ coloring of the graph $G$ with $|f_0| = \alpha(G)$. Suppose $(u, w) \notin f_0$.

**Case1.** $(u, w) \notin f_1$. As $f$ is a MPP $0-1$ coloring of the graph $G$, there is a $(w, w') \in E(G)$, such that $(w, w') \in f_0$. Consider the mapping $g : f_1 \cup (f_0 \setminus \{(w, w')\}) \cup \{(u, w)\} \to \{0, 1\}$ defined in the following way:

$$g(e) = \begin{cases} 0, & \text{if } e \in (f_0 \setminus \{(w, w')\}) \cup \{(u, w)\} \\ 1, & \text{if } e \in f_1. \end{cases}$$

It is clear that $g$ is a MPP $0-1$ coloring of the graph $G$, $(u, w) \in g_0$ and $|g_0| = |f_0| = \alpha(G)$.

**Case 2.** $(u, w) \in f_1$. As $f$ is a MPP $0-1$ coloring of the graph $G$, with $|f_0| = \alpha(G)$, then there is a $(w, w_1) \in f_0$. Consider the maximal alternating path $u, (u, w), w, (w, w_1), w_1, \ldots, w_{k-1}, (w_{k-1}, w_k), w_k$, where $k$ is odd, $\{(u, w), (w_1, w_2), \ldots, (w_{k-2}, w_{k-1})\} \subseteq f_1$ and $\{(w, w_1), (w_2, w_3), \ldots, (w_{k-1}, w_k)\} \subseteq f_0$. Define a mapping $g : f_0 \cup f_1 \to \{0, 1\}$ as follows:

$$g(e) = \begin{cases} f(e), & \text{if } e \notin \{(u, w), (w, w_1), \ldots, (w_{k-1}, w_k)\} \\ 1 - f(e), & \text{if } e \in \{(u, w), (w, w_1), \ldots, (w_{k-1}, w_k)\}. \end{cases}$$

Clearly, $g$ is a MPP $0-1$ coloring of the graph $G$ with $(u, w) \in g_0$ and $|g_0| = |f_0| = \alpha(G)$. The proof is complete.

**Lemma 2.** Let $G$ be a graph, $u \in V(G)$, $v \in V(G)$, $w \in V(G)$, $d_G(u) = d_G(v) = 1$, $(u, w) \in E(G)$, $(v, w) \in E(G)$. Then

(a) there is a MPP $0-1$ coloring $f$ of the graph $G$, such that $|f_0| = \alpha(G)$, $(u, w) \in f_0$ and $(v, w) \in f_1$;

(b) $\lambda(G) = 2 + \lambda(G \setminus \{u, v, w\})$, $\alpha(G) = 1 + \alpha(G \setminus \{u, v, w\})$.

**Proof.** (a) By **Lemma 1**, there is a MPP $0-1$ coloring $f$ of the graph $G$, such that $|f_0| = \alpha(G)$ and $(u, w) \in f_0$. Suppose $(v, w) \notin f_1$, then there is a $(w, w') \in E(G)$, such that $(w, w') \in f_1$. Consider a mapping $g : f_0 \cup (f_1 \setminus \{(w, w')\}) \cup \{(v, w)\} \to \{0, 1\}$ defined in the following way:

$$g(e) = \begin{cases} 1, & \text{if } e \in (f_1 \setminus \{(w, w')\}) \cup \{(v, w)\} \\ 0, & \text{if } e \in f_0. \end{cases}$$

Clearly, $g$ is a MPP $0-1$ coloring of the graph $G$ with $(u, w) \in g_0$, $(v, w) \in g_1$ and $|g_0| = \alpha(G)$.

(b) Let $w_1, \ldots, w_r$ be vertices of the graph $G$ such that $d_G(w) = r + 2$ $(r \geq 0)$, $u \notin \{w_1, \ldots, w_r\}$, $v \notin \{w_1, \ldots, w_r\}$, $(w, w_i) \in E(G)$ for $i = 1, \ldots, r$, and $f$ be a MPP $0-1$ coloring $f$ of the graph $G$, such that $|f_0| = \alpha(G)$, $(u, w) \in f_0$, $(v, w) \in f_1$. As $(w, w_i) \notin f_0 \cup f_1$ for $i = 1, \ldots, r$, we have

$\lambda(G) = \lambda(G \setminus \{(w, w_1), \ldots, (w, w_r)\}) = 2 + \lambda(G \setminus \{u, v, w\})$,
$\alpha(G) = \alpha(G \setminus \{(w, w_1), \ldots, (w, w_r)\}) = 1 + \alpha(G \setminus \{u, v, w\})$.

The proof is complete.

**Corollary.** Let $G$ be a graph, $U = \{u_0, u_1, u_2, u_3, u_4\}$ be a subset of the set of vertices of $G$ satisfying the conditions: $d_G(u_0) = d_G(u_4) = 1$, $d_G(u_1) = d_G(u_3) = 2$, $(u_{i-1}, u_i) \in E(G)$ for $i = 1, 2, 3, 4$. Then the following is true:

$$\lambda(G) = \lambda(G \setminus U) + 4, \quad \alpha(G) \geq 2 + \alpha(G \setminus U).$$

**Proof. Lemma 2** implies
$\lambda(G) = 2 + \lambda(G \setminus \{u_0, u_4\}) = \lambda(G \setminus U) + 4$, therefore $\alpha(G) \geq 2 + \alpha(G \setminus U)$.

**Theorem.** Let $G$ be a tree in which the distance between any two endpoints is even. Then the equality $\alpha(G) = \beta(G)$ holds.

**Proof.** Clearly, the statement of the theorem is true for the case $|E(G)| \leq 6$. Assume that it holds for trees with $|E(G)| \leq t - 1$, and let us prove that it will hold for the case $|E(G)| = t$, where $t \geq 7$.

**Case1.** There is a $U = \{u_0, u_1, u_2, u_3\} \subseteq V(G)$, such that $d_G(u_0) = 1$, $d_G(u_1) = d_G(u_2) = 2$, $(u_{i-1}, u_i) \in E(G)$ for $i = 1, 2, 3$. Set $G' = G\setminus\{u_0, u_1\}$. Clearly, $\beta(G) = \beta(G') + 1$. As $d_G(u_0) = 1$, $d_G(u_1) = 2$ and $d_{G\setminus\{u_0\}}(u_1) = 1$, $d_{G\setminus\{u_0\}}(u_2) = 2$, we have $\lambda(G) = 1 + \lambda(G\setminus\{u_0\}) = \lambda(G') + 2$, thus if $g$ is a MPP $0-1$ coloring of tree $G'$, such that $|g_0| = \alpha(G')$ and $(u_2, u_3) \in g_0$, then the mapping $f: g_0 \cup g_1 \cup \{(u_0, u_1), (u_1, u_2)\} \rightarrow \{0, 1\}$ defined as

$$f(e) = \begin{cases} g(e), & \text{if } e \notin \{(u_0, u_1), (u_1, u_2)\} \\ 1, & \text{if } e = (u_1, u_2) \\ 0, & \text{if } e = (u_0, u_1), \end{cases}$$

is a MPP $0-1$ coloring of the tree $G$, therefore $\alpha(G) \geq |f_0| = 1 + |g_0| = 1 + \alpha(G')$. As the distance between any two endpoints of $G'$ is even and $|E(G')| < t$, we have $\alpha(G') = \beta(G')$, therefore

$$\alpha(G) \geq 1 + \alpha(G') = 1 + \beta(G') = \beta(G), \text{ or } \alpha(G) = \beta(G).$$

**Case2.** There is a $U = \{u_0, u_1, u_2, u_3, u_4, u_5, u_6\} \subseteq V(G)$, such that $d_G(u_0) = d_G(u_4) = d_G(u_6) = 1$, $d_G(u_1) = d_G(u_3) = d_G(u_5) = 2$, $(u_{i-1}, u_i) \in E(G)$ for $i = 1, 2, 3, 4$, $(u_2, u_5) \in E(G)$, $(u_5, u_6) \in E(G)$. Set $G' = G\setminus\{u_5, u_6\}$. Clearly, $\beta(G) = \beta(G') + 1$. From **Corollary** follows that $\lambda(G) = \lambda(G\setminus\{u_0, u_1, u_2, u_3, u_4\}) + 4$, therefore $\lambda(G) = \lambda(G\setminus\{(u_2, u_5)\}) = \lambda(G') + 1$ and $\alpha(G) \geq 1 + \alpha(G')$. Note that the distance between any two endpoints of the tree $G'$ is even and $|E(G')| < t$, thus the equality $\alpha(G') = \beta(G')$ holds, and therefore

$$\alpha(G) \geq 1 + \alpha(G') = 1 + \beta(G') = \beta(G), \text{ or } \alpha(G) = \beta(G).$$

**Case3.** There is a $U = \{u_0, u_1, u_2\} \subseteq V(G)$, such that $d_G(u_0) = d_G(u_2) = 1$, $(u_{i-1}, u_i) \in E(G)$ for $i = 1, 2$. Let $D_1, \ldots, D_r$ be the connected components of $G\setminus U$. Clearly, $\beta(G) = 1 + \sum_{i=1}^{r} \beta(D_i)$. Note that for $i = 1, \ldots, r$ $D_i$ is a tree for which $|E(D_i)| < t$ and the distance between any two endpoints is even, thus $\alpha(D_i) = \beta(D_i)$, therefore, by **Lemma2**, we have

$$\alpha(G) = 1 + \alpha(G\setminus U) = 1 + \sum_{i=1}^{r} \alpha(D_i) = 1 + \sum_{i=1}^{r} \beta(D_i) = \beta(G).$$

**Case4.** There is a $U = \{u_0, u_1, u_2, u_3, u_4, u_5, u_6\} \subseteq V(G)$, such that $d_G(u_0) = d_G(u_6) = 1$, $d_G(u_1) = d_G(u_3) = d_G(u_5) = 2$, $d_G(u_2) = 3$, $(u_{i-1}, u_i) \in E(G)$ for $i = 1, 2, 3, 4, 6$, $(u_2, u_5) \in E(G)$. Set $G' = G\setminus\{u_0, u_1\}$. Clearly, $\beta(G) = \beta(G') + 1$. As $|E(G')| < t$ and the distance between any two endpoints of the tree $G'$ is even, the equality $\alpha(G') = \beta(G')$ holds.

**Lemma1** implies, that there is a MPP $0-1$ coloring $g$ of the tree $G\setminus\{u_0, u_1, u_2, u_5, u_6\}$ such that $(u_3, u_4) \in g_0$. Consider the mapping $f: g_0 \cup g_1 \cup \{(u_0, u_1), (u_2, u_3), (u_2, u_5), (u_5, u_6)\} \rightarrow \{0, 1\}$ defined as follows:

$$f(e) = \begin{cases} g(e), & \text{if } e \notin \{(u_0, u_1), (u_2, u_3), (u_2, u_5), (u_5, u_6)\} \\ 0, & \text{if } e \in \{(u_0, u_1), (u_2, u_5)\} \\ 1, & \text{if } e \in \{(u_2, u_3), (u_5, u_6)\}. \end{cases}$$

**Corollary** implies, that $f$ is a MPP $0-1$ coloring of the tree $G$, therefore $\lambda(G) = \lambda(G \setminus \{(u_1, u_2)\}) = \lambda(G') + 1$ and $\alpha(G) \geq 1 + \alpha(G') = \beta(G') + 1 = \beta(G)$, or $\alpha(G) = \beta(G)$.

**Case 5.** There is a $U = \{u_0, u_1, u_2, u_3, u_4, u_5, u_6, u_7\} \subseteq V(G)$, such that $d_G(u_0) = d_G(u_4) = d_G(u_6) = 1$, $d_G(u_1) = d_G(u_3) = 2$, $d_G(u_2) = d_G(u_5) = 3$, $(u_{i-1}, u_i) \in E(G)$ for $i = 1, 2, 3, 4, 6$, $(u_2, u_5) \in E(G)$, $(u_5, u_7) \in E(G)$. Set $G' = G \setminus \{u_0, u_1, u_2, u_3, u_4\}$. Note that $\beta(G) = \beta(G') + 2$. As $|E(G')| < t$ and the distance between any two endpoints of the tree $G'$ is even, the equality $\alpha(G') = \beta(G')$ holds. From **Corollary** we have

$$\alpha(G) \geq 2 + \alpha(G') = 2 + \beta(G') = \beta(G), \text{ or } \alpha(G) = \beta(G).$$

**Case 6.** There is a $U = \{u_0, u_1, u_2, u_3, u_4, u_5, u_6, u_7, u_8, u_9, u_{10}\} \subseteq V(G)$, such that $d_G(u_0) = d_G(u_4) = d_G(u_5) = d_G(u_9) = 1$, $d_G(u_1) = d_G(u_3) = d_G(u_6) = d_G(u_8) = 2$, $d_G(u_2) = d_G(u_7) = 3$, $(u_{i-1}, u_i) \in E(G)$ for $i = 1, 2, 3, 4, 6, 7, 8, 9$, $(u_2, u_{10}) \in E(G)$, $(u_7, u_{10}) \in E(G)$. Set $G' = G \setminus \{u_0, u_1, u_2, u_3, u_4\}$. Clearly $\beta(G) = \beta(G') + 2$. As $|E(G')| < t$ and the distance between any two endpoints of the tree $G'$ is even, the equality $\alpha(G') = \beta(G')$ holds, therefore from **Corollary** we have

$$\alpha(G) \geq 2 + \alpha(G') = 2 + \beta(G') = \beta(G), \text{ or } \alpha(G) = \beta(G).$$

As every tree $G$, in which the distance between any two endpoints is even, and $|E(G)| \geq 7$, satisfies at least one of the conditions of the six cases considered above, the proof of the **Theorem** is complete.

**Acknowledgement:** I would like to thank my Supervisor R. R. Kamalian for his constant attention to this work and for everything he has done for me.